\documentclass{pnastwomod}

\usepackage{graphicx}
\usepackage{graphics}
\usepackage{amssymb,amsfonts,amsmath}
\url{www.pnas.org/cgi/doi/...}
\copyrightyear{2009}
\issuedate{Issue Date}
\volume{Volume}
\issuenumber{Issue Number}
\usepackage{pdfpages}

\begin{document}

\title{A ratchet mechanism for amplification\\
in low-frequency mammalian hearing
}

\author{Tobias Reichenbach and A.~J.~Hudspeth\affil{1}{Howard Hughes Medical Institute and Laboratory of Sensory Neuroscience,
The Rockefeller University,
New York, New York 10065-6399, U.S.A.
}}

\contributor{Submitted to Proceedings of the National Academy of Sciences
of the United States of America}

\maketitle

\begin{article}

\begin{abstract}
The sensitivity and frequency selectivity of hearing result from tuned amplification by an active process in the mechanoreceptive hair cells.  In most vertebrates the active process stems from the active motility of hair bundles.  The mammalian cochlea exhibits an additional form of mechanical activity termed electromotility: its outer hair cells (OHCs) change length upon electrical stimulation.   The relative contributions of these two mechanisms to the active process in the mammalian inner ear is the subject of intense current debate. Here we show that active hair-bundle motility and electromotility can together implement an efficient mechanism for amplification that functions like a ratchet: sound-evoked forces acting on the basilar membrane are transmitted to the hair bundles whereas electromotility decouples active hair-bundle forces from the basilar membrane. This unidirectional coupling can extend the hearing range well below the resonant frequency of the basilar membrane. It thereby provides a concept for low-frequency hearing that accounts for a variety of unexplained experimental observations from the cochlear apex, including the shape and phase behavior of apical tuning curves,
their lack of significant nonlinearities, and the shape changes of threshold tuning 
curves of auditory nerve fibers along the cochlea. The ratchet mechanism constitutes a general design principle for implementing mechanical amplification in engineering applications.
\end{abstract}


\dropcap{T}he mammalian cochlea acts as a frequency analyzer in which high frequencies are detected at the organ's base and low frequencies at more apical positions. This frequency mapping is thought to be achieved by a position-dependent resonance of the elastic basilar membrane separating two fluid-filled compartments (Fig.~1A)~\cite{lighthill-1981-106,mammano-1993-93,shera-2004-5}. When sound evokes a pressure wave that displaces the basilar membrane, the resultant traveling wave gradually increases in amplitude as it progresses to the position where the basilar membrane's resonant frequency coincides with that of the stimulus. Aided by mechanical energy provided by the active process, the wave peaks at a characteristic place slightly before the resonant position and then declines sharply (Fig.~1B). This mechanism is termed critical-layer absorption~\cite{lighthill-1981-106}, for a wave cannot travel beyond its characteristic position on the basilar membrane, but peaks and dissipates most of its energy there. The mechanism displays scale invariance: different stimulation frequencies induce traveling waves that display a common, strongly asymmetric form upon rescaling of the amplitude and spatial coordinate~\cite{Siebert,Zweig}.

Two important aspects of the cochlea's mechanics remain problematical. First, the basilar membrane's resonant frequency apparently cannot span the entire range of audible frequencies. Experimental measurements of  basilar-membrane stiffness suggest that high-frequency  resonances are feasible but low-frequency ones are inaccessible~\cite{naidua-1998-124}. This result accords with the analysis of threshold tuning curves for auditory-nerve fibers~\cite{kiang-1986-171,Ulfendahl1997331,AndreiNTemchin11012008} and measurements of basilar-membrane displacement~\cite{Ulfendahl1997331,cooper-1995-82,zinn-2000-142,robles-2001-81}, both of which indicate that a peaked traveling wave occurs for high-frequency stimulation but not for low-frequency stimulation. Indeed, threshold tuning curves of auditory-nerve fibers~\cite{kiang-1986-171,Ulfendahl1997331,AndreiNTemchin11012008} show that high-frequency curves are scale-invariant and possess a sharp cutoff at frequencies above their characteristic frequencies, reflecting the mechanism of critical-layer absorption. However, low-frequency curves lack this sharp cutoff and, in contradiction of the expectation for critical-layer absorption, are instead characterized by an approximately symmetric shape around their characteristic frequencies. Recent experiments in the chinchilla have shown that the shape change between the tuning curves of high- and low-frequency fibers occurs between  two crossover frequencies of about 5 kHz and 1.5 kHz~\cite{Ulfendahl1997331,AndreiNTemchin11012008}. Measurements of basilar-membrane displacement yield similar conclusions. Experiments from the cochlear base confirm the existence at high frequencies  of peaked traveling waves that result in strongly asymmetric tuning curves and a pronounced nonlinearity at the characteristic frequencies~\cite{cooper-1992-63,ruggero-1997-101,Ulfendahl1997331,robles-2001-81}. Apical measurements of basilar-membrane displacement, however, produce symmetric tuning curves that lack a sharp cutoff for frequencies higher than the characteristic frequency~\cite{cooper-1995-82,khanna-1999-132,zinn-2000-142}. Moreover, only small nonlinearity has been measured~\cite{khanna-1999-132,cooper-1996-3,cooper-1995-82}, raising the question whether amplification occurs at the apex. All available experimental results therefore indicate that low-frequency hearing does not function through critical-layer absorption but must rely on another mechanism to achieve frequency selectivity.

The second key uncertainty is the nature of the active process that operates in the mammalian cochlea. In addition to amplifying weak signals, the active process produces increased frequency selectivity, compressive nonlinearity, and spontaneous otoacoustic emission. In the mammalian cochlea amplification is provided by specialized OHCs located in the organ of Corti along the basilar membrane (Fig.~2A). Each OHC displays two forms of motility. Like those in the hearing organs of other vertebrates~\cite{martin-1999-96,martin-2001-98-2}, the hair bundle of an OHC can produce mechanical force \cite{martin-1999-96,martin-2001-98-2,kennedy-2005-433,chan-2005-8,nadrowski-2004-101} (see~\cite{hudspeth-2009-59} for a review). But an OHC also exhibits electromotility: when its membrane potential changes as a result of sound-evoked hair-bundle deflection, the entire cell undergoes a length change owing to conformational rearrangement of the membrane protein prestin~\cite{zheng-2000-405,PeterDallos102006,ashmore-2008-88}. Direct measurements of the respective roles of the two forms of motility are complicated, for it is difficult to determine the micromechanical responses of the organ of Corti while the cochlea remains intact.

Here we have taken a theoretical approach to investigate a possible mechanism for amplification by the synergistic interplay of active hair-bundle motility and electromotility. We show that they can operate together to achieve low-frequency selectivity in the absence of basilar-membrane resonance and critical-layer absorption. The model advances theoretical understanding of cochlear mechanics in three ways: it reproduces previous theoretical results insofar as they coincide with experiments; it accounts naturally for a variety of unexplained findings from the cochlear apex; and it makes robust, experimentally testable predictions.\\

\section{Results}

\subsection{The Ratchet Mechanism}

Consider a transverse element of the cochlear partition comprising the basilar membrane, an inner hair cell (IHC) and three OHCs, and the overlying tectorial membrane (Fig.~2A). When a sound-evoked force displaces the basilar membrane upward, the resultant shearing motion between the tectorial membrane and the top of the OHCs, or reticular lamina, deflects the hair bundles in the positive direction. Active hair-bundle motility in the OHCs increases the amplitude of deflection. Without electromotility, this additional displacement would couple back to the basilar membrane and augment the movement there. If electromotility is adjusted such that the OHCs elongate just as much as the tectorial membrane and reticular lamina move upward, however, the basilar membrane does not experience the active force and thus undergoes no additional displacement (see also Movie S1).

\begin{figure}
%\centerline{\includegraphics[width=8.3cm]{fig1_naked_crop}}
\centerline{\includegraphics[scale=1]{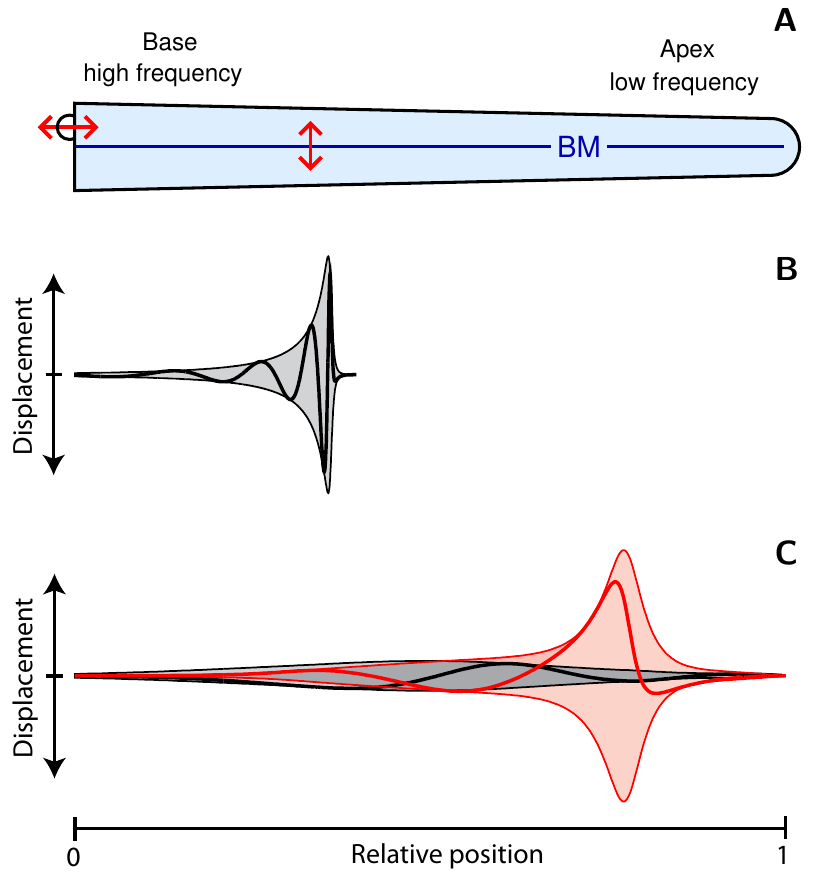}}
\caption{Principles of cochlear mechanics. (A) In a schematic diagram of the mammalian cochlea, the basilar membrane (BM) is displaced by sound stimuli acting on the stapes (top left). (B) In the classical theory of cochlear mechanics, sound evokes a pressure wave that causes a longitudinal traveling wave of basilar-membrane displacement (thick line). The motion of the basilar membrane and the displacements of the associated hair bundles are approximately equal. As the wave approaches the position where its frequency matches the basilar membrane's resonant frequency, the wave's amplitude (thin line) increases and its wavelength and velocity decline. The wave peaks at a characteristic place slightly before the resonant position and then declines sharply, yielding a strongly asymmetric envelope of the traveling wave (shading). Experiments confirm this behavior in the basal, high-frequency part of the cochlea. (C) We propose an alternative theory for the cochlea's mechanics at low frequencies. The basilar membrane near the cochlear apex does not resonate, but the traveling wave on the basilar membrane propagates along the entire cochlea without a strong variation in amplitude, wavelength, and velocity (black). However, the interplay of electromotility and active hair-bundle motility fosters an independent resonance of the complex formed by the hair bundles, reticular lamina, and tectorial membrane. The hair-bundle displacement (red) at the characteristic place can therefore exhibit an approximately symmetric peak, exceeding basilar-membrane motion by orders of magnitude.}\label{fig1}
\end{figure}
A mathematical formulation of this amplification mechanism clarifies its operation. Consider a two-mass model in which the two degrees of freedom are the motion of the basilar membrane, $X_\text{BM}$, and that of the complex formed by the hair bundles, reticular lamina, and tectorial membrane, $X_\text{HB}$ (Fig.~2B).  A sound stimulus of frequency $f$ produces an oscillating external force $F_\text{ext}(t)=\tilde{F}_\text{ext}e^{2\pi i f t} + c.c.$ acting on the basilar membrane, in which $c.c$ denotes the complex conjugate. The evoked  oscillations of the basilar membrane as well as  the hair-bundle complex occur predominantly  at the same frequency,  $X_\text{BM}=\tilde{X}_\text{BM}e^{2\pi i f t}+ c.c.$ and $X_\text{HB}=\tilde{X}_\text{HB}e^{2\pi i f t}+ c.c.$. Internal forces $F_\text{int}$  arise within hair bundles, where they provide  negative damping and introduce nonlinearities (Supporting Information).  The cell bodies of OHCs can be described as piezoelectric elements~\cite{dong-2002-82}. For small, physiologically relevant  motions their electrically evoked displacement $X_\text{EE}=\tilde{X}_\text{EE}e^{2\pi i f t}+ c.c.$ is proportional to the hair-bundle displacement, which triggers changes in the membrane potential, so  $\tilde{X}_\text{EE}=-\alpha\tilde{X}_\text{HB}+ c.c.$ with a complex mechanomotility coefficient $\alpha$~\cite{evans-1993-90}.  The hair-bundle and basilar-membrane displacements then depend linearly on  the forces:
\begin{equation}
 A\begin{pmatrix}  \tilde{X}_\text{HB} \\ \tilde{X}_\text{BM} \end{pmatrix} =\frac{1}{2\pi i  f}\begin{pmatrix}  \tilde{F}_\text{int} \\ \tilde{F}_\text{ext} \end{pmatrix} \,.
\label{matrixeq}
\end{equation}
The matrix  $A$ contains the impedances  $Z_\text{HB}$ and  $Z_\text{BM}$ of the hair-bundle complex and the basilar membrane as well as the coupling impedances $Z_\text{D}$  and $Z_\text{C}$  (Fig.~2B):
\begin{equation}
A=
\begin{pmatrix}
Z_\text{HB}+(1+\alpha)Z_\text{D}+Z_\text{C} & -Z_\text{D}-Z_\text{C}\\
-(1+\alpha)Z_\text{D}-Z_\text{C} & Z_\text{BM}+Z_\text{D}+Z_\text{C}
\end{pmatrix}\,.
\label{XHBXBM}
\end{equation}

\begin{figure}[t]
%\centerline{\includegraphics[width=7.2cm]{fig2_naked_crop}}
\centerline{\includegraphics[scale=1]{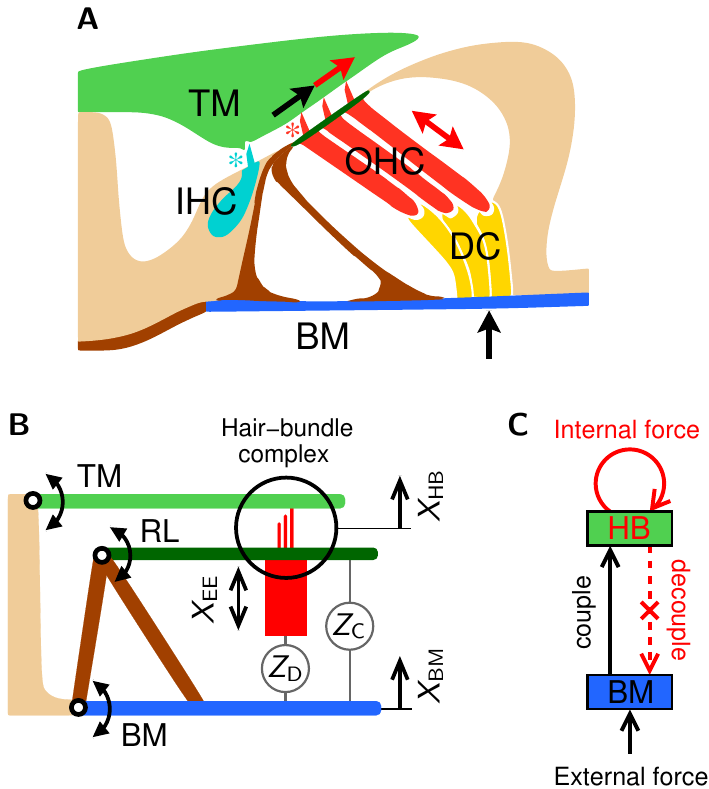}}
\caption{The ratchet mechanism. (A) The organ of Corti rests upon the basilar membrane (BM). Three OHCs are connected to Deiters' cells (DC), which together couple the basilar membrane to the reticular lamina (dark green, top of the OHCs) and through the hair bundles to the overlying tectorial membrane (TM). Sound-evoked external forces (black arrow) displace the basilar membrane, here upwards, and produce shearing (black arrow) of the hair bundles of OHCs (red asterisk) and the inner hair cell (IHC) (cyan asterisk). Two forms of motility underlie the active process: active hair-bundle motility (single-headed red arrow) and membrane-based electromotility (double-headed red arrow). (B) The two fundamental degrees of freedom are the basilar-membrane displacement  $X_\text{BM}$ and the displacement  $X_\text{HB}$ of the hair-bundle complex (circle), which comprises the hair bundles, reticular lamina (RL), and tectorial membrane. Coupling stems from the impedance $Z_\text{D}$  of the combined OHCs and Deiters' cells as well as the impedance $Z_\text{C}$   of the remaining organ of Corti. (C) In the ratchet mechanism displacements of the basilar membrane caused by external forces are communicated to the hair-bundle complex. Internal forces (red arrow) in the hair bundles increase the shearing motion, which decouples from basilar-membrane displacement through appropriate length changes of the OHCs (dotted red arrow). For an animated representation of the model, see Movie S1.}\label{fig2}
\end{figure}
A key feature of this relation is that  the matrix element $A_{21}$, which describes the coupling of the internal force to the basilar membrane, vanishes at a critical value $\alpha_*\equiv -1-Z_\text{C}/Z_\text{D}$.  At the same time, the coupling of the external force to the hair bundle, which is represented by element $A_{12}$, remains nonzero. For the critical value $\alpha_*$ the displacements become
\begin{eqnarray}
\tilde{X}_\text{HB}&=&\frac{1}{2\pi i f Z_\text{HB}}\tilde{F}_\text{int} +\frac{Z_\text{D}+Z_\text{C}}{2 \pi i f Z_\text{HB}(Z_\text{BM}+Z_\text{D}+Z_\text{C})}\tilde{F}_\text{ext} \,,\cr
\tilde{X}_\text{BM}&=&\frac{1}{2\pi i f(Z_\text{BM}+Z_\text{D}+Z_\text{C})}\tilde{F}_\text{ext}\,.
\label{XHBXBM*}
\end{eqnarray}
The hair-bundle displacement depends on both the internal and external forces, whereas the basilar membrane experiences only the external force. In other words, the sound-evoked external force on the basilar membrane is transmitted to the hair bundle, but the internal force from active hair-bundle motility does not feed back onto the motion of the basilar membrane (Fig.~2C). This symmetry-breaking mode has the characteristics of a ratchet in the sense that information flows unidirectionally: information applied as a force against the basilar membrane can be detected in the form of hair-bundle displacement, but force acting on the hair bundle cannot be detected at the basilar membrane.
The basilar-membrane displacement equals that occuring when the hair-bundle complex is fixed at $\tilde{X}_\text{HB}=0$ and electromotility is absent ($\alpha=0$). The ratchet mechanism therefore resembles an ideal operational amplifier that neither feeds back on nor draws energy from the input~\cite{Horowitz}.

\subsection{The Mechanomotility Coefficient $\alpha$}

\hspace*{1mm}The magnitude and phase of the critical value $\alpha_*$ depend on the coupling impedances  $Z_C$ and $Z_D$. Under realistic assumptions $|Z_C|$  is similar to or smaller than  $|Z_D|$  and $|\alpha_*|$ is therefore near unity. Such a value for the mechanomotility coefficient $\alpha$  has been measured experimentally for apical OHCs~\cite{evans-1993-90}. The phase of $\alpha$  is controlled by the complex network of ion channels that regulates the membrane potential depending on the hair-bundle displacement~\cite{evans-1993-90,sziklai-2004-261}. Although experiments on the phase of $\alpha$ are not available, theoretical considerations  confirm that OHCs have sufficient flexibility through ion-channel regulation to adjust the phase of $\alpha$ to a variety of values, and in particular to the phase required for $\alpha_*$ (Supporting Information).

\subsection{A Concept for Low-Frequency Hearing}

As its most important characteristic, the ratchet mechanism can explain frequency selectivity near the cochlear apex in the absence of a peaked traveling wave. The resonant frequency of the hair-bundle complex is determined solely by the impedance  $Z_\text{HB}$ and the internal forces, and is therefore defined by the properties of the hair bundles and tectorial membrane (Equation 3) ~\cite{JJZwislocki05111979,gummer-1996-93,HongxueCai04202004}. In particular, hair-bundle resonance can occur in the absence of basilar-membrane resonance. The ratchet mechanism therefore permits the resonant frequency of the hair bundles to follow a logarithmic law along the cochlea, whereas the resonant frequency of the basilar membrane remains significantly greater near the apex (Fig.~3B)~\cite{naidua-1998-124}.

Further evidence points to the occurrence of the ratchet mechanism at the cochlear apex but not at the base. First, the basilar membrane at the base is narrow and presumably tuned to the characteristic frequencies at which auditory-nerve fibers are most sensitive (Fig.~3B). Amplification of basilar-membrane motion there is feasible and need not be avoided. Indeed, experiments have demonstrated a strong compressive nonlinearity of basilar-membrane motion at high frequencies. However, the apex exhibits a wider basilar membrane that experiences stronger viscous forces and whose resonant frequencies deviate from the characteristic frequencies~\cite{naidua-1998-124}. Amplification of basilar-membrane motion near the apex would therefore be highly inefficient. The absence of a strong compressive nonlinearity in experimental measurements of basilar-membrane displacement in the apical region confirms the lack of basilar-membrane amplification there. Next, the membrane time constant of OHCs restricts electromotility's ability to work on a cycle-by-cycle basis to frequencies below a few kilohertz~\cite{JSantos-Sacchi05011992,housley-1992-448}, disabling the ratchet mechanism for higher frequencies. It is noteworthy that the ion channels of apical OHCs differ significantly from those of basal OHCs~\cite{engel-2006-143} and that apical OHCs are considerably longer than basal ones and can accordingly produce greater length changes~\cite{ashmore-2008-88}.

\subsection{A Unified Model for Cochlear Mechanics}

We have quantified these considerations in a one-dimensional model of the cochlea (Supporting Information). Studies of threshold tuning curves of auditory-nerve fibers in the chinchilla have delineated three distinct regimes that are separated by two crossover frequencies, a high frequency $f_\text{H}\approx 5$~kHz and a low frequency $f_\text{L}\approx 1.5$~kHz (Fig.~3)~\cite{AndreiNTemchin11012008}. In our model for the basal regime I, above $f_\text{H}$, electromotility is negligible owing to the membrane time constant; each segment of the basilar membrane is tuned to its characteristic frequency. In the transitional regime II, between $f_\text{H}$ and $f_\text{L}$, unidirectional coupling provided by electromotility occurs but every segment of the basilar membrane still resonates at its characteristic frequency. In the apical regime III, below $f_\text{L}$, electromotility combines with active hair-bundle motility to implement the ratchet mechanism for amplification. The resonant frequency of the basilar membrane is therefore of minor importance and remains approximately constant at a value significantly above the range of characteristic frequencies.

\begin{figure}[t]
%\centerline{\includegraphics[width=8.7cm]{fig3_naked_crop}}
\centerline{\includegraphics[scale=1]{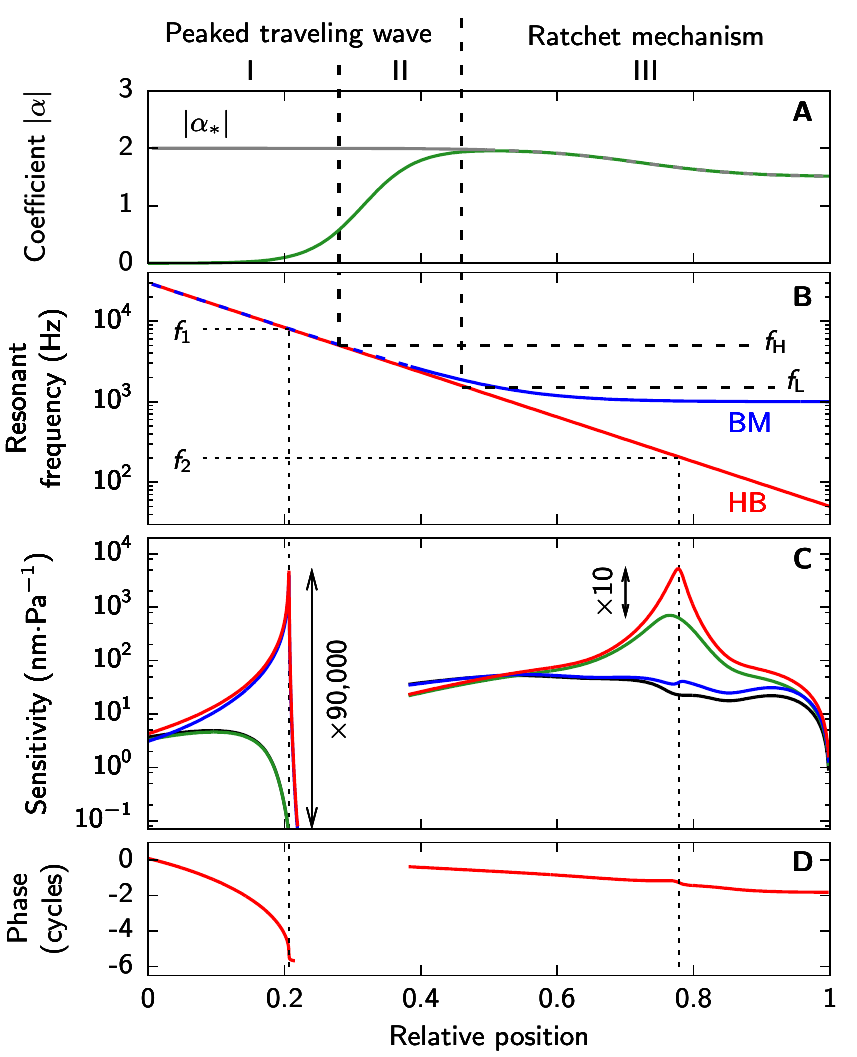}}
\caption{Cochlear model. (A) The ratchet mechanism operates when the mechanomotility coefficient $\alpha$ (green) coincides with the critical value $\alpha_*$ (grey). Although electromotility is negligible to the basal side of $f_\text{H}$, it underlies the ratchet mechanism apical to the position of $f_\text{L}$. (B) The resonant frequency of the hair-bundle complex (HB, red) agrees with that of the basilar membrane (BM, blue) only for frequencies above $f_\text{L}$. (C) A high-frequency sound stimulus ($f_1=8$ kHz)  induces  a traveling wave that peaks in the basal region. The displacements of the hair bundles (red) coincide with that of the basilar membrane (blue). Elimination of active hair-bundle motility  decreases the sensitivity by a factor of $90,000$ (green, hair bundles; black, basilar membrane) 
 indicative of a strong nonlinearity. A low-frequency stimulus ($f_2=200$ Hz)  triggers a traveling wave that does not peak on the basilar membrane, but the hair-bundle displacement exhibits a resonance enabled by the ratchet mechanism (same color code). Without active hair-bundle motion the hair-bundle displacement decreases by a factor of only  ten indicative of a
weak nonlinearity.  (D) The phase of the basilar-membrane displacement for $f_1$ has a strongly increasing slope near the resonant position and thus shows a wave traveling to the resonant position but not beyond. For $f_2$ the slope of the phase remains almost constant, corresponding to a wave traveling beyond the characteristic place.
}\label{fig3}
\end{figure}
In accordance with the theory of critical-layer absorption~\cite{lighthill-1981-106}, in regime I the traveling wave produced by a pure-tone stimulus peaks at the characteristic place and sharply decreases beyond that point (Fig.~3C). The characteristic behavior of critical-layer absorption -- decrease of the traveling wave's speed and wavelength near the characteristic place -- appears in the phase behavior as an increase of the slope (Fig.~3D). Active hair-bundle motility amplifies the peak displacement. A strong compressive nonlinearity arises because amplification by active hair-bundle motility both increases the hair-bundle and basilar-membrane motion per unit pressure difference and enhances the amplitude of the pressure wave itself (Fig.~3C).  The combination of the two effects yields a compressive nonlinearity that extends over a significantly broader range of sound intensities than the nonlinearity in hair-bundle motion itself.

In regime II electromotility influences the micromechanics and causes hair-bundle displacement to exceed basilar-membrane displacement. The theory of critical-layer absorption still applies. The basilar-membrane tuning curves, phase behavior, and nonlinearity consequently parallel those in regime I.

In regime III the ratchet mechanism leads to very different behavior. Sound evokes a low-frequency pressure wave that traverses the entire cochlea, evoking only a small basilar-membrane displacement throughout, for the resonant frequency of the basilar membrane is higher everywhere (Fig.~3B). At the characteristic place, however, the hair bundles exhibit an independent resonance amplified through active hair-bundle motility and facilitated by the unidirectional coupling provided by electromotility. The hair-bundle displacement there can exceed the basilar-membrane response by orders of magnitude (Fig.~3C). In further contrast to the theory of critical-layer absorption, the traveling wave does not slow and its wavelength does not vanish at the characteristic place, for the basilar membrane does not resonate. This becomes apparent in the behavior of the traveling wave's phase, whose slope remains low and nearly constant across the characteristic place (Fig.~3D). The phase behavior also confirms that the wave reaches the helicotrema. Because amplification through the ratchet mechanism does not act on the basilar membrane, the latter  exhibits approximately linear behavior (Fig.~3C). The pressure wave is therefore unaffected by the active process. Hair-bundle displacement is amplified and exhibits a moderate compressive nonlinearity (Fig.~3C).

\subsection{Threshold Tuning Curves of Auditory-Nerve Fibers}

\begin{figure}
%\centerline{\includegraphics[width=8.7cm]{fig5_naked_crop}}
\centerline{\includegraphics[scale=1]{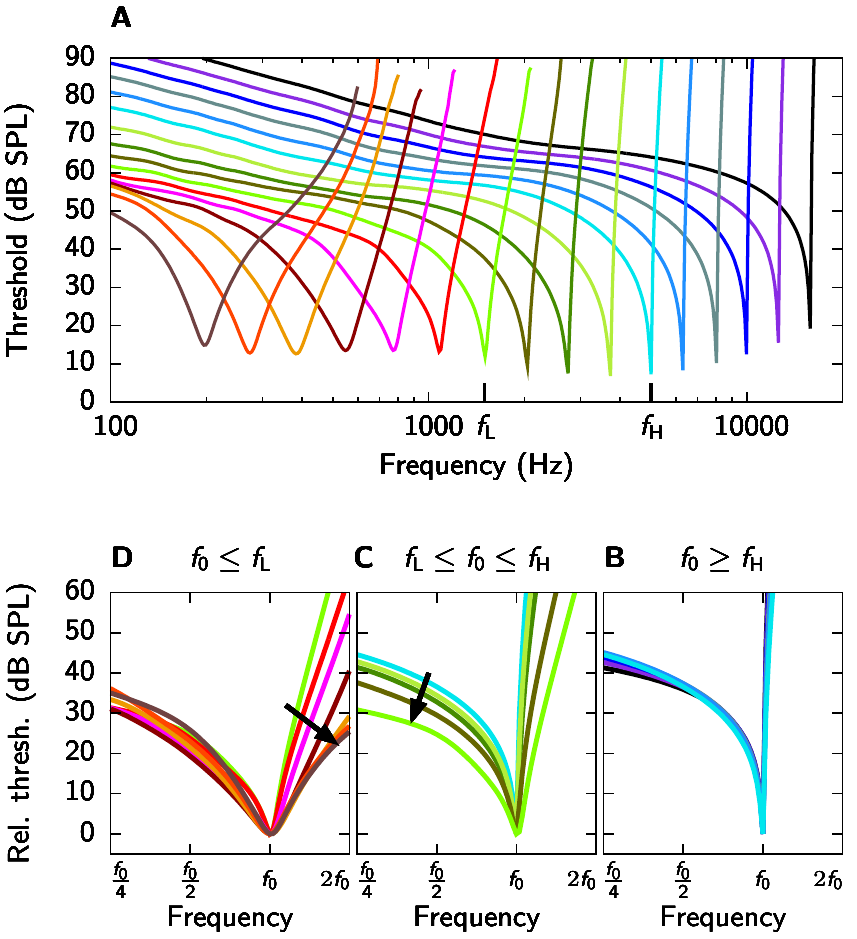}}
\caption{Threshold tuning curves of auditory-nerve fibers. (A) The tuning curve at each position along the cochlea has a characteristic frequency $f_0$ corresponding to the resonant frequency of the hair-bundle complex. When tuning curves are rescaled such that the frequency is measured in octaves relative to the characteristic frequency $f_0$ and the threshold is measured relative to that at the characteristic frequency, characteristic shape changes are seen to occur between curves of different characteristic frequencies. (B) Tuning curves for high characteristic frequencies, above $f_\text{H}$, fall onto a universal curve that exhibits the strongly asymmetric form and high-frequency cutoff characteristic of a peaked traveling wave. (C) As the characteristic frequency declines from $f_\text{H}$ to $f_\text{L}$, the left limb falls (arrow), indicating the emerging influence of electromotility and the ratchet mechanism. (D) As the characteristic frequency diminishes below $f_\text{L}$, the right limb falls steeply (arrow), pointing to the breakdown of the peaked-wave mechanism and the dominance of ratchet amplification.}\label{fig4}
\end{figure}
Strong support for the proposed model comes from studies of threshold tuning curves for auditory-nerve fibers ~\cite{kiang-1986-171,Ulfendahl1997331,AndreiNTemchin11012008}. Recent measurements have shown characteristic shape changes occurring around the crossover frequencies $f_\text{H}$ and $f_\text{L}$~\cite{AndreiNTemchin11012008}. To compare these data to our model, we have computed tuning curves and -- despite their complexity and the simplicity of our model -- found striking agreement with the measurements (Fig.~4). High-frequency fibers, those tuned above $f_\text{H}$, display the strongly asymmetric shape characteristic of critical-layer absorption. Moreover, they fall onto a universal curve when the frequency and threshold are measured relative to the values at the characteristic frequency. This accords with experimental findings and the scaling symmetry of the peaked traveling-wave mechanism~\cite{Siebert,Zweig,AndreiNTemchin11012008}. For intermediate characteristic frequencies between $f_\text{L}$ and $f_\text{H}$, the scaling law is violated as the ratchet mechanism starts to influence the micromechanics, causing the left limb of the threshold tuning curves to fall as the characteristic frequency decreases~\cite{AndreiNTemchin11012008}. A second violation of the scaling arises for characteristic frequencies below $f_\text{L}$, for which the right limb falls. Also observed experimentally ~\cite{AndreiNTemchin11012008}, this second violation of scaling indicates a breakdown of critical-layer absorption, which predicts a steep increase for threshold tuning curves at frequencies above the characteristic frequency.

\section{Discussion}

The classical theory of cochlear mechanics assumes that the basilar membrane resonates at a characteristic place for each frequency in the auditory range. The resulting mechanism of critical-layer absorption is characterized by ({\it i}) strongly asymmetric, scale-invariant tuning curves with steep high-frequency cutoffs, ({\it ii}) an increasing slope of the traveling wave's phase upon approaching the characteristic place, ({\it iii}) basilar-membrane displacement similar to hair-bundle displacement, and ({\it iv}) pronounced compressive nonlinearity at the characteristic frequency. This picture is consistent with diverse experimental findings from the cochlear base, including direct measurements of basilar-membrane motion and studies of threshold tuning curves for auditory-nerve fibers. However, as summarized in the Introduction, experimental results from the cochlea's apex are in qualitative disagreement with the first, second, and fourth characteristics.

Here we have proposed a concept for low-frequency hearing that employs a ratchet mechanism involving the interplay of active hair-bundle motility and electromotility. Whereas sound-evoked forces displace the basilar membrane and the hair bundles, the active forces within the hair bundle can decouple from the basilar membrane through appropriate elongation and contraction of OHCs. This mechanism allows hair bundles to resonate independently of the basilar membrane and thus explains how the ear's hearing range can extend to values well below the resonant frequency of the basilar membrane. We have shown that amplification by the ratchet mechanism leads to characteristic behavior that is distinct from that associated with critical-layer absorption: it exhibits ({\it i$^\prime$}) approximately symmetric tuning curves that display no sharp high-frequency cutoff, ({\it ii}$^\prime$) a constant slope of the traveling wave's phase across the characteristic place, ({\it iii}$^\prime$) tuned hair-bundle displacement that exceeds the untuned basilar-membrane displacement by orders of magnitude at resonance, and ({\it iv}$^\prime$) approximately linear basilar-membrane displacement and only moderate compressive nonlinearity in hair-bundle motion at the characteristic frequency. 

\subsection{Comparison to Experimental Results}

The proposed model yields the classical theory of critical-layer absorption for high-frequency sounds (Figs.~1B,~3 and~4A). Experiments in the base have confirmed ({\it i}) the asymmetric form of tuning curves~\cite{cooper-1992-63,ruggero-1997-101}, ({\it ii}) the increasing slope of the phase~\cite{rhode-1971-49,nuttall-1996-99}, and ({\it iv}) the strong compressive nonlinearity~\cite{cooper-1992-63,nuttall-1996-99,ruggero-1997-101}. Owing to difficulties in accessing the motion of the organ of Corti at the base, the relation ({\it iii}) of hair-bundle motion to basilar-membrane motion has not yet been measured there.

Because the membrane time constant of OHCs is expected to limit the operation of electromotility to low frequencies~\cite{JSantos-Sacchi05011992,housley-1992-448}, we have chosen in our model to consider only active hair-bundle motility in the basal part of the cochlea. However, studies in prestin knockout mice suggest that electromotility is necessary for high-frequency amplification~\cite{mellado-2008-11,Dallos2008333}. Electromotility near the base may adjust the operating point of the hair-bundle motor on a time-scale slower than the period of oscillation or provide a fast electrical signal in the form of membrane-potential change upon mechanical stimulation.

Our theory accounts for a number of unexplained results from the cochlear apex. Experiments have measured ({\it i}$^\prime$) approximately symmetric tuning curves without sharp high-frequency cutoffs of tectorial-membrane and therefore hair-bundle displacement~\cite{cooper-1995-82,khanna-1999-132,zinn-2000-142}. In our model the approximate symmetry arises naturally, for it reflects the resonance of the hair bundles, reticular lamina, and tectorial membrane independently of resonance by the basilar membrane (Figs.~1C and 3C,D). Experiments have consistently reported ({\it ii}$^\prime$) a constant phase slope across the characteristic place~\cite{cooper-1995-82,cooper-1996-2,zinn-2000-142}. This feature emerges in our model (Fig.~3D) because the basilar membrane is untuned at low frequencies and the pressure wave therefore reaches the helicotrema. There remains a controversy about the existence of nonlinearity at the cochlear apex (reviewed in Ref.~\cite{robles-2001-81}): although some investigators have found no nonlinearites~\cite{khanna-1999-132}, others have reported a small compressive~\cite{cooper-1995-82,cooper-1996-3} or expansive nonlinearity~\cite{zinn-2000-142}. However, all agree about ({\it iv}$^\prime$) the absence of a strong compressive nonlinearity.

The cochlear apex allows observation of different parts of the organ of Corti -- such as the tectorial membrane and Hensen's cells as well as the basilar membrane -- and thus permits tests of characteristic ({\it iii}$^\prime$) of our model. Experiments on a temporal-bone preparation~\cite{khanna-1999-132,ITER-1989-467}  have shown that vibrations of the reticular lamina and tectorial membrane  are orders of magnitude larger than the basilar-membrane displacement at the characteristic place. These findings are in perfect agreement with our theory. However, the in vitro experiments have been criticized for an uncertainty about which constituents of the organ of Corti were measured~\cite{robles-2001-81} and are in contradiction with studies using other preparations~\cite{cooper-1995-82,cooper-1996-3}. Definite conclusions must therefore be postponed until further experimental results become available.

Perhaps the most reliable comparison of our model to experimental measurements comes through tuning curves of auditory-nerve fibers, which yield information about cochlear mechanics that is least disturbed by experimental intervention and represent responses from the whole length of the cochlea. These tuning curves exhibit progressive shape changes from ({\it i}) the strongly asymmetric form typical of high-frequency fibers to ({\it i}$^\prime$) the nearly symmetric form found for low-frequency fibers~\cite{kiang-1986-171,Ulfendahl1997331,AndreiNTemchin11012008}. Our modeled responses of tuning curves of auditory-nerve fibers exhibit the same shape changes (Fig.~4), thus providing a conceptual explanation for the three distinct cochlear regimes.

\subsection{Experimentally Testable Predictions}

Our theory yields experimentally testable predictions based on the characteristics ({\it i}$^\prime$)-({\it iv}$^\prime$). As discussed above, ({\it i}$^\prime$) the approximately symmetric tuning curves as well as ({\it ii}$^\prime$) the constant phase slope have already been observed. In contrast, characteristic ({\it iii}$^\prime$), which predicts hair-bundle motion that exceeds the basilar-membrane motion at the characteristic frequency by orders of magnitude, remains controversial. Future experiments are therefore required to test this prediction. Such studies should also determine whether, as our theory implies, the complex formed by the hair bundles, reticular lamina, and tectorial membrane exhibits a resonance independent of the basilar membrane, whose response is untuned. A further test is feasible through experimental studies of the nonlinear behavior at the apex, for which our theory predicts ({\it iv}$^\prime$) a moderate compressive nonlinearity in hair-bundle motion, and therefore in tectorial-membrane motion, but no more than a weak nonlinearity in the basilar membrane's response. Finally, and beyond the theory presented in this article, the ratchet mechanism should allow  for traveling waves in the tectorial membrane~\cite{RoozbehGhaffari10162007}  that are unidirectionally coupled to basilar-membrane waves.

\subsection{The Ratchet Principle}

The ratchet mechanism constitutes a general design principle for mechanical amplification whereby the output does not feed back onto the input. In this way it represents a mechanical analogue of the operational amplifier from electrical engineering~\cite{Horowitz}. Although the current technology of signal detectors such as microphones relies on electrical amplification, mechanical amplification could improve signal-to-noise ratios and thereby greatly advance sensitivity and detection of weak signals. The ratchet mechanism thus opens a path for implementing controlled mechanical amplification in engineering applications.

\begin{materials}
\section{Hydrodynamics}  By combining continuity equations and fluid-momentum equations, we describe the cochlea's hydrodynamics with the partial differential equation
  \begin{equation}
 \rho \partial_t^2X_\text{BM}+\Lambda\partial_t X_\text{BM}= \frac{1}{2L^2}\partial_r\left( h\partial_r p \right)\,.
 \label{eq:cochlea}
 \end{equation}
 Here $\rho$ denotes  the density of liquid in the cochlea,  $L$ the length of the cochlea, $h$  the height of the scalae,  and $p$ and $X_\text{BM}$ respectively the pressure across  and the displacement of the basilar membrane at position $r$ and time $t$. The coefficient $\Lambda$ accounts for friction due to  fluid motion. Position is measured in units of the cochlear length, such that $r=0$ corresponds to the basal and $r=1$ to the apical end. The pressure translates into an external force acting on the basilar membrane and yields a displacement that we compute employing the model of Fig.~2B (Supporting Information). Temporal Fourier transformation yields an ordinary differential equation that we solve numerically with the shooting method in Mathematica 7 (Wolfram Research).  We apply two boundary conditions.  First, $p=p_0$ at $r=0$: a sound-evoked pressure $p_0$ acts at the stapes.  And second, $p=0$ at $r=1$: because the two scalae communicate at the helicotrema, the pressure difference between them vanishes at the apical end of the cochlea. 
 
\section{Tuning Curves of  Auditory-Nerve Fibers} Tuning curves of auditory-nerve fibers are computed by assuming that the hearing threshold corresponds to a root-mean-square deflection of the hair bundles of IHCs by $0.3$ nm. These bundles are thought to be coupled by fluid motion to the shearing between the reticular lamina and tectorial membrane, and thus to the displacement of the hair bundles of OHCs (Supporting Information).
\end{materials}

\begin{acknowledgments}
We thank S. Leibler for discussion, the two reviewers for valuable suggestions, and the members of our research group for comments on the manuscript. This research was supported by grant DC00241 from the National Institutes of Health and by a fellowship to T. R. from the Alexander von Humboldt Foundation. A. J. H. is an Investigator of Howard Hughes Medical Institute.
\end{acknowledgments}

%\vspace*{-5mm}

%\bibliographystyle{unsrt}

%\bibliography{/Users/tobias/UNI/literature/lit_hearing}



\section*{Supplementary material}

# Supporting Information
# A Ratchet Mechanism for Amplification in Low-Frequency Mammalian Hearing

Tobias Reichenbach and A. J. Hudspeth

*Howard Hughes Medical Institute and Laboratory of Sensory Neuroscience, The Rockefeller University, New York, New York 10065-6399, U.S.A.*

Here we further describe our modeling approach. We provide a detailed description of the two-mass model for the organ of Corti and discuss the energy balance. We then turn to the hydrodynamics of the cochlea. Because our goal is to demonstrate a principle rather than to describe the cochlea in full detail, we focus on a basic, one-dimensional model. We conclude with a model for the regulation of the OHCs' membrane potential which demonstrates how the critical mechanomotility coefficient $\alpha_*$ can result.

## Micromechanics of the organ of Corti

To describe the micromechanics of the organ of Corti we employ a model with two degrees of freedom: the motion of the basilar membrane and the motion of the hair-bundle complex (Figure 2B). The impedance $Z_{\text{BM}}$ of the basilar membrane results from a mass $m_{\text{BM}}$, viscous damping $\lambda_{\text{BM}}$, and stiffness $K_{\text{BM}}$. Similarly, the impedance of the complex formed by the hair bundles, reticular lamina, and tectorial membrane involves mass $m_{\text{TM}}$, viscous damping $\lambda_{\text{HB}}$, and stiffness $K_{\text{HB}}$.

The motions of the basilar membrane and of the hair-bundle complex are coupled through the impedance $Z_{\text{C}}$ of the organ of Corti, which possesses viscous and elastic contributions $\lambda_{\text{C}}$ and $K_{\text{C}}$. Further coupling arises through the combined OHCs and Deiters' cells (Fig. S1). An OHC can be described as a piezoelectric element [1]. Its length change $\delta L$ depends both on a change $\delta V$ in the membrane potential as well as on the applied force $F$. Considering oscillatory motions at angular frequency $\omega = 2\pi f$ we can write $\delta L = \widetilde{\delta L} e^{i\omega t} + c.c.$, $\delta V = \widetilde{\delta V} e^{i\omega t} + c.c.$, and $F = \widetilde{F} e^{i\omega t} + c.c.$ in which $c.c$ denotes the complex conjugate. The length change then follows as

$$\widetilde{\delta L} = -c\widetilde{\delta V} + c_2 \widetilde{F} \tag{5}$$

with coefficients $c$, $c_2$. We define $\tilde{X}_{\text{EE}} = -c\widetilde{\delta V}$ as the part of the length change that results from voltage changes alone. The length change $c_2 \widetilde{F}$ results from the OHC's viscoelasticity $Z_{\text{OHC}} = -i\omega^{-1} c_2^{-1}$ in series with $X_{\text{EE}}$ (Fig. S1). It lies in series with the Deiters' cell which can be described as another viscoleastic element $Z_{\text{DC}}$. The two viscoelastic elements $Z_{\text{OHC}}$ and $Z_{\text{DC}}$ in series combine into the viscoelastic element

$$Z_{\text{D}} = (Z_{\text{OHC}}^{-1} + Z_{\text{DC}}^{-1})^{-1} \tag{6}$$

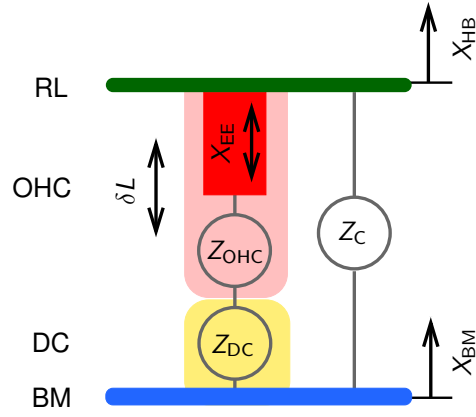

Figure S1: Coupling between the reticular lamina (RL) and the basilar membrane (BM). The coupling arises from OHCs in series with Deiters' cells (DC) and an impedance $Z_C$ from the remaining organ of Corti. See the text for a description of the coupled OHCs and Deiters' cells.

such that the schematic of Fig. 2B results. We denote by $K_\text{D}$ the elastic component and by $\lambda_\text{D}$ the viscous part of the impedance $Z_\text{D}$. The equations of motion for hair-bundle and basilar-membrane displacement $X_\text{HB}$ and $X_\text{BM}$ then read

$$m_\text{TM}\partial_t^2 X_\text{HB} + \lambda_\text{HB}\partial_t X_\text{HB} + K_\text{HB} X_\text{HB} + (K_D + \lambda_D \partial_t)(X_\text{HB} - X_\text{EE} - X_\text{BM}) + (K_C + \lambda_C \partial_t)(X_\text{HB} - X_\text{BM}) = F_\text{int},$$
$$m_\text{BM}\partial_t^2 X_\text{BM} + \lambda_\text{BM}\partial_t X_\text{BM} + K_\text{BM} X_\text{BM} - (K_D + \lambda_D \partial_t)(X_\text{HB} - X_\text{EE} - X_\text{BM}) - (K_C + \lambda_C \partial_t)(X_\text{HB} - X_\text{BM}) = F_\text{ext}. \quad (7)$$

A sound stimulus at angular frequency $\omega$ provides an external force $F_\text{ext}$ on the basilar membrane whose dependence on time $t$ may be written as $F_\text{ext}(t) = \tilde{F}_\text{ext} e^{i\omega t} + c.c.$ with the Fourier coefficient $\tilde{F}_\text{ext}$. The evoked oscillations of the basilar membrane, $X_\text{BM}$, and of the hair-bundle complex, $X_\text{HB}$, occur dominantly at the same frequency $f$; additional frequencies may arise from nonlinear effects owing to internal forces in the hair bundle. We consider the corresponding Fourier coefficients $\tilde{X}_\text{BM}$ and $\tilde{X}_\text{HB}$. The electrically evoked length change $\tilde{X}_\text{EE}$ of OHCs depends linearly on the hair-bundle displacement, for the membrane-potential change depends linearly on hair-bundle motion (see Supplementary Section on the electrical regulation of the OHC membrane potential): $\tilde{X}_\text{EE} = -\alpha \tilde{X}_\text{BM}$ with a complex, frequency-dependent mechanomotility coefficient $\alpha$. Supplementary Equation (7) then yields Equations (1) and (2) with

$$\begin{aligned} Z_\text{HB} &= i\omega m_\text{TM} + \lambda_\text{HB} - i\omega^{-1} K_\text{HB}, \\ Z_\text{BM} &= i\omega m_\text{BM} + \lambda_\text{BM} - i\omega^{-1} K_\text{BM}, \\ Z_\text{C} &= \lambda_\text{C} - i\omega^{-1} K_\text{C}, \\ Z_\text{D} &= \lambda_\text{D} - i\omega^{-1} K_\text{D}. \end{aligned} \quad (8)$$

The forces within the hair bundle depend on its displacement; their effect is to counter viscous damping as well as to influence the resonant frequency of the hair-bundle complex. We may decompose these forces into linear and

nonlinear parts:

$$\tilde{F}_{\text{int}} = i\omega Z_{\text{HB}}^{\text{act}} \tilde{X}_{\text{HB}} + \tilde{F}_{\text{int}}^{\text{nonlin}}(\tilde{X}_{\text{HB}}) \,. \tag{9}$$

Here the linear term dominates the dynamics for small displacements ensuing from weak sound stimuli and describes the fully active scenario. The nonlinear term connects the active to the passive case that arises for strong sound stimuli.

At the critical value $\alpha_*$, the displacements are given by Equations (3), which can be rewritten using Supplementary Equation (9) as

$$\begin{aligned}
\tilde{X}_{\text{HB}} &= \frac{1}{i\omega(Z_{\text{HB}} - Z_{\text{HB}}^{\text{act}})} \tilde{F}_{\text{int}}^{\text{nonlin}}(\tilde{X}_{\text{HB}}) + \frac{Z_{\text{D}} + Z_{\text{C}}}{i\omega(Z_{\text{HB}} - Z_{\text{HB}}^{\text{act}})(Z_{\text{BM}} + Z_{\text{D}} + Z_{\text{C}})} \tilde{F}_{\text{ext}} \,, \\
\tilde{X}_{\text{BM}} &= \frac{1}{i\omega(Z_{\text{BM}} + Z_{\text{D}} + Z_{\text{C}})} \tilde{F}_{\text{ext}} \,.
\end{aligned} \tag{10}$$

Resonance in hair-bundle motion occurs at the frequency for which $Z_{\text{HB}} - Z_{\text{HB}}^{\text{act}} = 0$. Active hair-bundle motility makes this situation possible. As discussed in the main text, this resonance condition is independent of the properties of the basilar membrane.

## Cochlear hydrodynamics and traveling waves

We now incorporate the description for the micromechanics of the organ of Corti into a model for the whole cochlea. In the simplest form, the cochlea is considered as one-dimensional, exhibiting a slow pressure wave. Combining continuity equations and fluid-momentum equations, this pressure wave obeys the partial differential equation

$$\rho \partial_t^2 X_{\text{BM}} + \Lambda \partial_t X_{\text{BM}} = \frac{1}{2L^2} \partial_r (h \partial_r p) \,. \tag{11}$$

Here $\rho$ denotes the density of liquid in the cochlea, $L$ the length of the cochlea, $h$ the height of the scalae, and $p$ and $X_{\text{BM}}$ respectively the pressure across and the displacement of the basilar membrane at position $r$ and time $t$. The coefficient $\Lambda$ accounts for friction due to fluid motion. Position is measured in units of the cochlear length, such that $r = 0$ corresponds to the basal and $r = 1$ to the apical end. We apply two boundary conditions. First, $p = p_0$ at $r = 0$: a sound-evoked pressure $p_0$ acts at the stapes. And second, $p = 0$ at $r = 1$: because the two scalae communicate at the helicotrema, the pressure difference between them vanishes at the apical end of the cochlea.

Considering the Fourier coefficients of angular frequency $\omega$, we obtain

$$-\omega^2 \rho \tilde{X}_{\text{BM}} + i\omega \Lambda \tilde{X}_{\text{BM}} = \frac{1}{2L^2} \partial_r (h \partial_r \tilde{p}) \,. \tag{12}$$

The basilar-membrane displacement $\tilde{X}_{\text{BM}}$ and pressure $\tilde{p}$ are connected through Equation (1), with $\tilde{p} = \tilde{F}_{\text{ext}}/A_{\text{BM}}$. $A_{\text{BM}}$ denotes the area of a transverse strip of the basilar membrane that has the width of one cell, about 8 $\mu$m; all parameters employed in the two-mass model refer to a transverse segment of the cochlear partition of that width. We solve Supplementary Equation (12) numerically with the shooting method in Mathematica 7 (Wolfram Research).

The height $h$ of the scalae varies from about 1 mm at the base to about 200 $\mu$m at the apex. Measurements of the fluid pressure in the basal region indicate, however, that the penetration depth of the wave is significantly smaller [2]. In our one-dimensional simulations, we therefore use an effective height of 100 $\mu$m.

Concerning the map of characteristic frequencies $f_0(r)$ along the cochlea, we consider the chinchilla's hearing range from a maximal frequency of about 30 kHz at the extreme base to a minimum of about 50 Hz at the extreme apex [3]. We assume that the resonant frequency of the basilar membrane cannot cover this whole range and is instead given by

$$f_{\rm BM}(r) = \sqrt{f_{\rm BM,apex}^2 + [1 - (f_{\rm BM,apex}/f_{\rm max})^2] f_0(r)^2}\,. \tag{13}$$

Here $f_{\rm max} = 30$ kHz denotes the highest characteristic frequency and $f_{\rm BM,apex} = 1$ kHz the lowest resonant frequency that the basilar membrane exhibits. $f_{\rm BM}$ coincides with $f_0$ for high frequencies but deviates for frequencies below $f_{\rm L}$ and approaches $f_{\rm BM,apex}$ in the apical region of the cochlea (Figure 3B).

The resonant frequency $f_{\rm BM}$ of the basilar membrane is set by its mass and stiffness according to $f_{\rm BM} = (2\pi)^{-1}\sqrt{K_{\rm BM}/m_{\rm BM}}$. We consider $K_{\rm BM}$ to vary proportional to $f_{\rm BM}$ and $m_{\rm BM}$ to vary inversely: $K_{\rm BM}(r) = K_{\rm BM}^{\rm max} f_{\rm BM}(r)/f_{\rm max}$ and $m_{\rm BM}(r) = K_{\rm BM}(r)/[2\pi f_{\rm BM}(r)]^2$, with $K_{\rm BM}^{\rm max} = 2$ N·m$^{-1}$. The mass $m_{\rm TM}$ of the tectorial membrane is assumed to follow the same spatial variation as the basilar membrane mass, albeit with a smaller value: $m_{\rm TM} = m_{\rm BM}/5$. The increase in the basilar-membrane and tectorial-membrane mass from base to apex presumably reflects the increasing size of the organ of Corti towards the apex.

For the variation of the mechanomotility coefficient $\alpha$ along the cochlea, and its dependence on the stimulus frequency $f$, we make the ansatz

$$\alpha(r,f) = \frac{f_c^4}{f_c^4 + f_0(r)^4} \times \frac{[\delta f(r) f_0]^2}{[f^2 - f_0(r)^2]^2 + [\delta f f_0(r)]^2} \times \alpha_*\,. \tag{14}$$

The first factor accounts for the high-frequency cutoff above which the membrane potential, and thus electromotility, can no longer follow hair-bundle displacement on a cycle-by-cycle basis; we use $f_c = 4$ kHz. To yield the ratchet mechanism at lower frequencies, we choose $\alpha$ to be $\alpha_*$ at the natural frequency. The second factor describes how the response changes when the frequency $f$ of stimulation deviates from the natural frequency $f_0$; we assume that $\alpha$ is close to $\alpha_*$ as long as both frequencies are similar, but otherwise declines in magnitude. The width of the corresponding curve is determined by the parameter $\delta$, which we set at $\delta = 2$. The mechanomotility coefficient $\alpha(r, f = f_0)$ is shown in Figure 3A.

The linear part of the active hair-bundle force counters viscous damping and provides a vanishing impedance at the natural frequency. Using Equations (1), (2), and Supplementary Equation (9), this translates into

$$Z_{\rm HB} - Z_{\rm HB}^{\rm act}\Big|_{f=f_0} = -\frac{Z_{\rm BM}[(1+\alpha)Z_{\rm D} + Z_{\rm C}]}{Z_{\rm BM} + Z_{\rm C} + Z_{\rm D}}\bigg|_{f=f_0}. \tag{15}$$

We assume that the imaginary part of $Z_{\rm HB}$ is tuned to the natural frequency $f_0$ such that the active contribution possesses only a real part corresponding to negative damping, $Z_{\rm HB}^{\rm act} \equiv \lambda_{\rm HB}^{\rm act}$. Note that in the ratchet mechanism, when $\alpha = \alpha_*$, the right-hand side of Supplementary Equation (15) vanishes, so active hair-bundle forces need to overcome only damping associated with hair-bundle motion.

| Parameter | Description | Value | Reference |
|---|---|---|---|
| $L$ | length of the cochlea | 20 mm | [4] |
| $\Lambda$ | fluid friction | $5 \times 10^5$ N·s·m$^{-4}$ | |
| $w_{\mathrm{BM}}(r)$ | width of the basilar membrane | $(50 + 200r)$ $\mu$m | [5] |
| $A_{\mathrm{BM}}(r)$ | area of the basilar-membrane strip | $w_{\mathrm{BM}}(r) \times 8$ $\mu$m | |
| $\lambda_{\mathrm{HB}}$ | friction of the hair-bundle complex | 1 $\mu$N·s·m$^{-1}$ | [6] |
| $K_{\mathrm{IHB}}$ | stiffness of the hair bundles of IHCs | 2 mN·m$^{-1}$ | [7] |
| $\lambda_{\mathrm{IHB}}$ | friction of the hair bundles of IHCs | 1 $\mu$N·s·m$^{-1}$ | [6] |
| $\lambda_{\mathrm{BM}}(r)$ | friction of the basilar membrane | $0.03 \times w_{\mathrm{BM}}(r)$ N·s·m$^{-2}$ | |
| $K_{\mathrm{C}}$ | stiffness of the organ of Corti | 10 mN·m$^{-1}$ | |
| $\lambda_{\mathrm{C}}$ | friction of the organ of Corti | 10 $\mu$N·s·m$^{-1}$ | |
| $K_{\mathrm{D}}$ | stiffness of OHCs and Deiters' cells | 20 mN·m$^{-1}$ | [8] |
| $\lambda_{\mathrm{D}}$ | friction of OHCs and Deiters' cells | 10 $\mu$N·s·m$^{-1}$ | |

S1: Parameters used in the cochlear model.

When $\alpha$ and active hair-bundle forces obey Supplementary Equations (14) and (15) and nonlinearities from the hair-bundle forces are ignored, the hair-bundle displacement diverges at resonance. In the actual cochlea, noise as well as nonlinearities supervene and lead to a large but finite displacement. In our simulations, we incorporate this effect by considering values for $\alpha$ and $Z_{\mathrm{HB}}^{\mathrm{act}}$ that deviate by 1% from their ideal values given by Supplementary Equations (14) and (15).

### Tuning curves of auditory-nerve fibers

Tuning curves of auditory-nerve fibers are computed by assuming that the hearing threshold corresponds to a root-mean-square deflection of the hair bundles of IHCs by 0.3 nm. These bundles are thought to be coupled by fluid motion to the shearing between the reticular lamina and tectorial membrane, and thus to the displacement of the hair bundles of OHCs. Assuming that the displacement $X_{\mathrm{IHB}}$ of the hair bundles of inner hair cells is dependent on friction $\lambda_{\mathrm{IHB}}$ and elasticity $K_{\mathrm{IHB}}$, it couples to the OHCs' bundle motion through the relation

$$\lambda_{\mathrm{IHB}} \partial_t (X_{\mathrm{IHB}} - X_{\mathrm{HB}}) + K_{\mathrm{IHB}} X_{\mathrm{IHB}} = 0. \tag{16}$$

Upon Fourier transformation, we obtain

$$\tilde{X}_{\mathrm{IHB}} = \frac{i\omega \lambda_{\mathrm{IHB}}}{i\omega \lambda_{\mathrm{IHB}} + K_{\mathrm{IHB}}} \tilde{X}_{\mathrm{HB}}. \tag{17}$$

For high frequencies, viscous coupling is strong and the hair-bundle displacements of IHCs and OHCs coincide. For low frequencies, the coupling decreases and the hair-bundle motion of IHCs remains smaller than that of OHCs. This effect underlies the rising thresholds at low frequencies seen on the left limbs of high-frequency tuning curves (Figure 5) as well as in experimental measurements.

The value of 0 dB SPL is defined by a root-mean-square sound-pressure stimulus of 20 $\mu$Pa. Because this

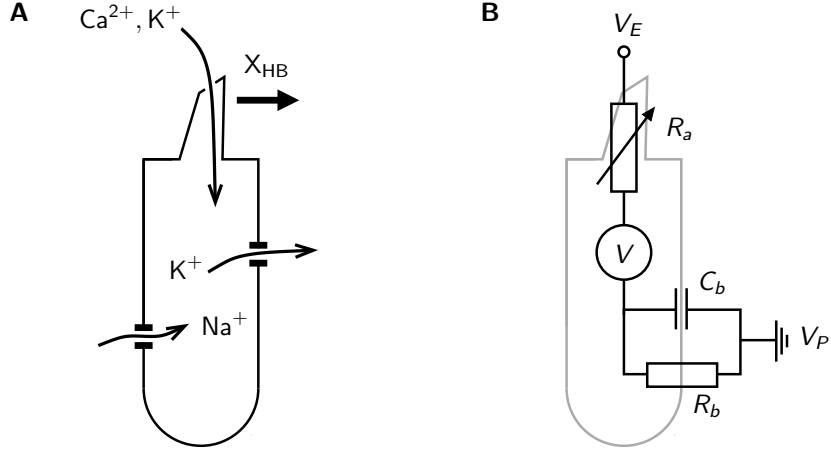

Figure S2: Regulation of the OHC membrane potential. (A) Hair-bundle deflection opens ion channels, $K^+$ and $Ca^{2+}$ flow into the cell. The cell membrane contains various additional ion channels that are regulated by the membrane potential and the $Ca^{2+}$ concentration. In our model, we consider voltage-regulated outward $K^+$ channels and inward $Na^+$ channels. (B) An equivalent circuit of an OHC with a variable resistance $R_a$ controlled by hair-bundle displacement, a membrane capacitance $C_b$, and a basoleteral membrane conductance $R_b$.

external pressure is enhanced by the middle ear before acting on the stapes, we assume an increase by a factor of 20 (26 dB) [9].

If they have not been detailed in the text above, the parameters of the two-mass model as well as the one-dimensional descriptions of the cochlea and hair bundle dynamics of IHCs are given in Table S1.

### Electrical regulation of the OHC membrane potential

Deflection $X_{\mathrm{HB}}$ of the OHCs hair bundles evokes a change $\delta V \equiv V - V_0$ in the membrane potential from its resting value $V_0$, which leads to a length change $X_{\mathrm{EE}} = -c\delta V$ (see Supplementary Equation 5 and below) with a coefficient $c$ of about 20 $\mu$m·V$^{-1}$ [1]. The ratchet mechanism relies on a critical value $\alpha_* = -1 - Z_{\mathrm{C}}/Z_{\mathrm{D}}$ of the ratio $\alpha = -X_{\mathrm{EE}}/X_{\mathrm{HB}}$ between hair-bundle displacement and cell length change. While its exact value depends on the coupling impedances $Z_{\mathrm{C}}$ and $Z_{\mathrm{D}}$, the real part of $\alpha_*$ is negative. A typical situation $Z_{\mathrm{C}} = Z_{\mathrm{D}}$ leads to the critical value $\alpha_* = -2$. Here we show how such a value of $\alpha_*$ can emerge within a basic model for the regulation of the OHCs membrane potential. The model relies on voltage-regulated outward $K^+$ channels [10] that counteract the current through the membrane capacitance, as well as a voltage-regulated inward sodium current [11; 12] that can lead to hyperpolarization upon positive hair-bundle deflection.

The apical surface of an OHC, including the hair bundle, is bathed in endolymph at a potential $V_E$ of about 80 mV (Figure S2). The basolateral part is surrounded by perilymph at ground potential $V_P$. The electrical behavior can be represented by a simple circuit described by an equation for current conservation:

$$I_{R_a} = I_{R_b} + I_{C_b}. \tag{18}$$

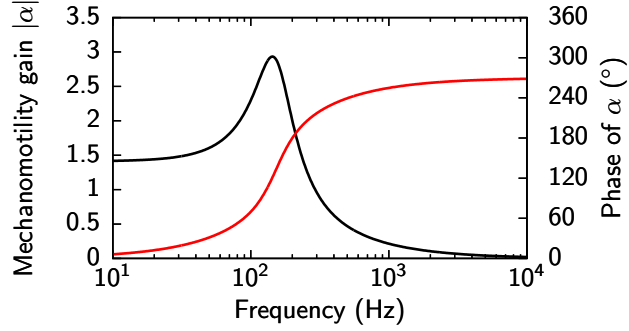

Figure S3: The modeled mechanomotility coefficient $\alpha$. We show the dependence of the magnitude (black) and phase (red) of $\alpha$ on the sound frequency. For $f_0 = 200$ Hz $\alpha = -2$ emerges which represents a typical value for the critical $\alpha_*$. The magnitude of $\alpha$ is strongly attenuated for higher frequencies. By varying the voltage dependency of the basal K$^+$ and Na$^+$ channels the critical value $\alpha_*$ can emerge at different frequencies $f_0$ below a few kilohertz.

Here $I_{R_a}$ denotes the current through the resistance $R_a$ and $I_{R_b}$ and $I_{C_b}$ denote the currents through $R_b$ and $C_b$, respectively. We assume that $R_a$ results from the conductivity $g_a$ of the mechanotransduction channels in the hair bundle; the current is mainly carried by K$^+$ ions [13]. The conductance of the basolateral cell membrane stems from voltage-regulated K$^+$ channels (conductance $g_b^{\mathrm{K}}$) as well as voltage-regulated Na$^+$-channels [11; 12] (conductance $g_b^{\mathrm{Na}}$). We obtain

$$g_a(V_E - V) = C_b \partial_t V + g_b^{\mathrm{K}}(V - V_P) - g_b^{\mathrm{Na}}(V_{\mathrm{Na}} - V) \tag{19}$$

in which $V_{\mathrm{Na}}$ denotes the Na$^+$ reversal potential and $C_b$ the cell membrane's capacitance. The potential $V_0$ inside the cell at vanishing hair-bundle displacement $X_{\mathrm{HB}} = 0$ follows as

$$V_0 = \frac{g_a^{(0)} V_E + g_b^{\mathrm{Na}\,(0)} V_{\mathrm{Na}} + g_b^{\mathrm{K}\,(0)} V_P}{g_a^{(0)} + g_b^{\mathrm{Na}\,(0)} + g_b^{\mathrm{K}\,(0)}} \tag{20}$$

in which the superscript (0) denotes the respective conductances at vanishing hair-bundle displacement.

Hair-bundle displacement $X_{\mathrm{HB}}$ changes the conductance $g_a$ of the cell's apical membrane. To linear oder we obtain

$$g_a = g_a^{(0)} + \left.\frac{\partial g_a}{\partial X_{\mathrm{HB}}}\right|_{(0)} X_{\mathrm{HB}} . \tag{21}$$

The altered apical conductance evokes a changes $\delta V = V - V_0$ in the membrane potential which influences the basal K$^+$ and Na$^+$ conductances. The K$^+$ channels are voltage-regulated at an activation time $\tau$. The dynamics of the change $\delta g_b^{\mathrm{K}} = g_b^{\mathrm{K}} - g_b^{\mathrm{K}\,(0)}$ can be described by

$$\tau \partial_t \delta g_b^{\mathrm{K}} = \left.\frac{\partial g_b^{\mathrm{K}}}{\partial V}\right|_{(0)} \delta V - \delta g_b^{\mathrm{K}} \tag{22}$$

| Parameter | Description | Value | Reference |
|---|---|---|---|
| $V_E$ | potential of the endolymph | 80 mV | [14] |
| $V_P$ | potential of the perilymph | $-100$ mV | [14] |
| $V_{\text{Na}}$ | Na$^+$ reversal potential | 22 mV | [11] |
| $g_a^{K^+}$ | apical conductivity of K$^+$ channels | 5 nS | [15] |
| $g_b^K$ | basoleteral K$^+$ conductivity | 10 nS | [10] |
| $C_b$ | membrane capacitance | 40 pF | [10] |
| $\partial g_a^{K^+}/\partial X_{\text{HB}}$ | dependence of apical K$^+$ conductivity on $X_{\text{HB}}$ | $5 \times 10^{-3}$ S·m$^{-1}$ | [16] |
| $\tau$ | activation time of voltage-regulated K$^+$ channels | 1 ms | [10] |

S2: Parameters for modeling the regulation of the OHCs membrane potential.

which results in

$$\widetilde{\delta g}_b^K = \frac{1}{1+i\omega\tau}\left.\frac{\partial g_b^K}{\partial V}\right|_{(0)}\widetilde{\delta V} \qquad (23)$$

for oscillatory stimuli $X_{\text{HB}} = \tilde{X}_{\text{HB}}e^{i\omega t} + c.c.$ at angular frequency $\omega$. The Na$^+$ channels are rapidly activated; to linear order their conductance follows as

$$g_b^{\text{Na}} = g_b^{\text{Na (0)}} + \left.\frac{\partial g_b^{\text{Na}}}{\partial V}\right|_{(0)}\delta V. \qquad (24)$$

Linearization of Supplementary Equation (19) yields

$$\left[i\omega C_b + g_a + g_b^K + g_b^{\text{Na}} + \frac{1-i\omega\tau}{1+\omega^2\tau^2}\left.\frac{\partial g_b^K}{\partial V}\right|_{(0)}(V-V_P) - \left.\frac{\partial g_b^{\text{Na}}}{\partial V}\right|_{(0)}(V_{\text{Na}}-V)\right]\widetilde{\delta V} = \left.\frac{\partial g_a}{\partial X_{\text{HB}}}\right|_{(0)}\tilde{X}_{\text{HB}}. \qquad (25)$$

The mechanomotility coefficient $\alpha$ follows as $\alpha = c\widetilde{\delta V}/\tilde{X}_{\text{HB}}$. From Supplementary Equation (25) we note that delayed voltage-regulated K$^+$ channels can reduce the effect of the membrane capacitance. A large enough voltage dependence of the inward Na$^+$ channels can then yield hyperpolarization of the cell upon positive deflection of the hair bundle, as requested for the critical value $\alpha_*$.

Figure S3 shows the resulting magnitude and phase of $\alpha$ for typical parameters for apical OHCs (Table S2) in dependence on the frequency of hair-bundle stimulation. The values $\partial g_b^K/\partial V = 1.36\ \mu S \cdot V^{-1}$ and $\partial g_b^{\text{Na}}/\partial V = 1.85\ \mu S \cdot V^{-1}$ have been chosen to yield the critical value $\alpha_* = -2$ at the frequency $f_0 = 200$ Hz and lie in the range of measured values [10; 11].



\end{article}

\end{document}